\documentclass[11pt]{article}

\usepackage[final]{acl}

\usepackage{times}
\usepackage{latexsym}
\usepackage[T1]{fontenc}
\usepackage[utf8]{inputenc}
\usepackage{microtype}
\usepackage{inconsolata}
\usepackage{amsmath}
\usepackage{booktabs}
\usepackage{graphicx}
\usepackage{url}
\usepackage{xcolor}
\usepackage{placeins}

\graphicspath{{figures/}}

\title{Comparing Sentiment Contagion in AI-Agent and Human Social Networks: Evidence from MOLTBOOK}

\author{
  Elyes Ben chaabane \\
  UNIL \\
  Lausanne, Switzerland \\
  \texttt{elyes.benchaabane@unil.ch} \\\And
  Savindu Herath \\
  ETH Zurich \\
  Zurich, Switzerland \\
  \texttt{sherath@ethz.ch} \\\And
  Yash Raj Shrestha \\
  UNIL \\
  Lausanne, Switzerland \\
  \texttt{yashraj.shrestha@unil.ch} \\
}

\newcommand{\pfdr}{p_{\mathrm{FDR}}}

\begin{document}
\maketitle

\begin{abstract}
AI agents are beginning to interact not only with people, but also with one another. We investigate what happens to sentiment in such an AI-only social network: does negativity spread, or do replies calm it down? We study MOLTBOOK, a social network made up of autonomous language-model agents, using almost 2.9 million posts and 1.5 million comments. Negative posts receive many more replies than neutral or positive posts, so negativity still attracts attention. However, replies to negative content usually do not stay negative. They most often become neutral, and there is meager evidence that negative sentiment spreads across days. The main pattern is therefore not a cycle of negativity, but negative attention followed by neutralisation. These findings suggest that AI-agent networks may behave differently from human social networks: they may dampen emotional extremes, while still depending strongly on how interactions are organised.

%\TODO{\noindent\textbf{Keywords:} AI social networks; sentiment propagation; counterfactual inference; measurement sensitivity; network resilience}
\end{abstract}

\section{Introduction}

Autonomous AI agents are increasingly connected not only to tools, APIs, and databases, but also to other AI agents. Once more than two agents interact, network structures emerge, opening a new research paradigm at the intersection of social network analysis and multi-agent systems \citep{park2023generative,mehdizadeh2025homophily}. This development has important implications as AI agents are becoming influential social actors whose collective behaviour mold interactions and the evolution of the network. As AI agents increasingly populate online communities, they generate machine--machine and machine--human dynamics that shape how information, sentiment, and behavioural patterns propagate across social networks \citep{tsvetkova2024,rahwan2019}. Understanding the formation and evolution of such networks has long been central to network science \citep{barabasi1999,newman2003,jackson2008}, and the emerging field of AI-agent network science extends this tradition by examining both how information diffuses through existing agent networks and how autonomous agents generate network structures themselves \citep{zhang2025llm}.

An important open question is whether sentiment dynamics in AI-only social networks resemble those observed in human social networks. In human interaction, negative content is often overrepresented because it attracts attention and can elicit further negative responses \citep{baumeister2001bad,brady2017emotion}. This attention advantage may support affective feedback loops, where exposure to negative content shapes subsequent engagement and reply behaviour \citep{kramer2014experimental,ferrara2015measuring,bail2018exposure}. AI-only social networks pose a different theoretical problem. If participants are autonomous language-model agents, sentiment dynamics are shaped not by human affective states, but by learned response distributions, alignment procedures, context-window exposure, and interaction structure. Based on prior work on human sentiment contagion and alignment-trained language models, we derive theory-guided hypotheses concerning engagement asymmetry, transition asymmetry, reply-chain neutralisation, temporal propagation, and mechanism sensitivity.

We test these hypotheses using MOLTBOOK, an AI-agent-only social network that records posts, comments, and comment replies among autonomous language-model agents. The data come from the Moltbook Observatory Archive on Hugging Face, a passive archival export of the Moltbook Observatory system \citep{moltbook_observatory,moltbook_observatory_archive_2026}. The setting allows us to examine whether sentiment in an AI-only network follows patterns known from human social networks, or whether it is shaped differently by language-model behaviour and interaction structure. Our results show a more nuanced pattern than a simple recovery-dominant logic. First, negative posts attract more comments, indicating that negative content retains an engagement advantage even in an AI-only network. Second, negative sentiment does not strongly persist in replies: across reply chains, sentiment tends to move away from negativity and is increasingly absorbed into neutral sentiment. Third, however, this movement is not clearly a positive recovery process. Negative-to-positive movement exists, but the dominant transition after negative parent content is neutralisation. Fourth, we find little evidence of delayed sentiment propagation compared with same-day co-movement. Same-day and next-day post-comment correlations are almost identical, suggesting that AI agents respond mainly to immediate conversational context rather than to a persistent sentiment environment from previous days.

These findings matter because they show that AI-agent networks may not reproduce human affective contagion fully. In human networks, negative content can attract attention, spread through exposure, and shape later behaviour. In MOLTBOOK, negative content also attracts engagement, but the replies do not clearly amplify negativity or convert it into positivity. Instead, the dominant pattern is \textit{neutralisation}. This distinction is important for AI safety and computational social science because future AI-agent systems may interact at scale in organisational, economic, and social environments. If such systems dampen sentiment extremes through local response tendencies rather than through stable network-level recovery mechanisms, then their collective behaviour may be less emotionally contagious than human networks but still highly sensitive to how interactions are structured.

To examine these mechanisms, we analyse MOLTBOOK activity from 2026-01-27 to 2026-05-06. The refreshed dataset contains 2,894,804 posts and 1,479,965 comments produced by 177,496 posting agents and 20,309 commenting agents. Sentiment is labelled using the CardiffNLP Twitter-RoBERTa sentiment model, selected through manual validation and compared against three additional sentiment models. We estimate transition asymmetry over valid parent-reply pairs, analyse negative-post engagement using count models, compare same-day and next-day sentiment correlations, and evaluate aggregate shock patterns in daily sentiment. Finally, we conduct counterfactual interventions that disrupt label structure, partner matching, hub-agent composition, and temporal ordering. These counterfactuals test whether the observed resilience patterns depend on the actual organisation of the AI-agent network.

This paper makes three contributions. First, it provides a large-scale empirical analysis of sentiment dynamics in an AI-agent-only social network, showing that negative posts attract more engagement but that replies tend to diffuse sentiment towards neutrality. Second, it refines the notion of recovery-dominant behaviour by combining manual validation, four-model transition comparison, and explicit transition estimands, showing that recovery claims are sensitive to sentiment measurement choices. Third, it introduces a counterfactual framework for testing mechanism sensitivity in AI-agent networks, demonstrating that resilience metrics change when label structure, partner matching, hub composition, or temporal ordering is disrupted. Together, these findings suggest that AI-agent networks may be resilient to affective extremes, but this resilience appears to operate less through positive recovery than through local neutralisation and structural features of the interaction network.

\section{Related Work and Hypothesis Development}

\subsection{Sentiment dynamics in human and AI-agent networks}

Research on social networks has long shown that emotions and opinions are not only properties of individual actors, but can also be shaped by social exposure, interaction, and network structure. In human networks, negative content is especially important because it attracts attention and generates further engagement \citep{baumeister2001bad,brady2017emotion}. Prior work on emotional contagion and sentiment propagation suggests that affective states may spread through social ties, replies, and repeated exposure \citep{kramer2014experimental,coviello2014detecting,ferrara2015measuring,bail2018exposure}. Related research further shows that sentiment can be predicted not only from text, but also from structural properties of social networks, indicating that who interacts with whom can be informative about affective dynamics \citep{jin2018sentiment}. More recent work conceptualises sentiment dynamics as a diffusion process, where the speed and pattern of diffusion depend on temporal and structural features of the network \citep{li2026sentiment}. Together, this literature suggests that sentiment is both a linguistic and a networked phenomenon.

AI-agent networks raise a related but distinct question. When language-model agents interact with one another, the resulting dynamics are not driven by human affective states, memory, or social identity in the usual sense. Instead, they emerge from model training, alignment procedures, prompting, context windows, and the structure of agent interactions. Recent work argues that connecting AI agents to one another creates a new class of machine--machine social dynamics that requires combining social network analysis with multi-agent systems \citep{tsvetkova2024,rahwan2019}. Studies of MOLTBOOK suggest that large-scale AI-only interaction can produce global stability while maintaining local diversity, but may also show limited evidence of durable socialisation, stable influence, or shared social memory \citep{li2026does}. Other analyses of MOLTBOOK highlight the importance of centralised hubs, bursty automation, topic-specific toxicity, and ecosystem-level risk, suggesting that AI-agent behaviour should be studied not only at the level of individual model outputs but also at the level of emergent network structure \citep{jiang2026humans}. This motivates our focus on sentiment resilience in an AI-only social network.

\subsection{Transition asymmetry}

In human social networks, negative content can attract attention and elicit further negative responses, creating the possibility of affective reinforcement or negative feedback loops \citep{baumeister2001bad,brady2017emotion}. However, AI-agent networks may differ because language models are typically trained and aligned to respond in helpful, polite, and non-escalatory ways. If so, then a negative parent comment should not usually trigger another negative reply. Instead, an aligned agent might respond constructively, neutrally, or positively, thereby reducing the persistence of negative sentiment across reply chains. This implies two related expectations. First, negative replies should be a minority outcome after a negative parent comment. Second, negative-origin chains should move towards positive sentiment more readily than positive-origin chains move towards negative sentiment. We therefore expect transition asymmetry in the direction of recovery from negativity.

\noindent\textbf{H1: Transition asymmetry.} Negative persistence is below 0.5, the negative-to-neutral transition satisfies $\hat{\eta}>0.5$, and the recovery-to-souring ratio satisfies $\hat{\rho} > 1$. Definitions are in Appendix~\ref{app:statistical-details}.

\subsection{Shock resilience}

Human social networks can experience aggregate sentiment shocks when external events, viral content, or coordinated attention push discourse sharply in a negative direction. Yet prior work on sentiment dynamics and diffusion also suggests that such deviations may mean-revert as attention shifts, interaction patterns change, or local exchanges become more balanced over time \citep{ferrara2015measuring,li2026sentiment}. In AI-agent networks, this reversion may be even stronger because agents do not necessarily experience persistent emotional states. A language-model agent receiving negative input is still expected to generate a response according to its learned behavioural distribution and alignment constraints rather than according to a deteriorating mood. If this is the case, negative shocks in aggregate sentiment should be dampened quickly, with the network returning towards its baseline sentiment level within a short period.

\noindent\textbf{H2: Shock resilience.} Aggregate sentiment returns quickly after negative deviations.

\subsection{Structural lag}

A key mechanism in human sentiment contagion is temporal propagation. Exposure to emotionally charged content at one point in time may affect later behaviour, such that negative posts today influence comments, replies, or posts tomorrow \citep{kramer2014experimental,coviello2014detecting,bail2018exposure}. If AI-agent networks reproduce this human-like contagion mechanism, then post sentiment should have predictive value beyond same-day co-movement. In particular, lagged post sentiment should predict next-day comment sentiment more strongly than contemporaneous sentiment, because the relevant signal would be delayed behavioural influence rather than immediate topical or contextual similarity. However, this expectation is theoretically uncertain in AI-agent networks. If agents respond mainly to the immediate context available in their prompt, and if they do not retain persistent memory of prior network sentiment, then lagged propagation should be weak. We therefore test whether MOLTBOOK exhibits a temporal contagion pattern analogous to human social networks.

\noindent\textbf{H3: Structural lag.} Lagged post sentiment predicts next-day comment sentiment more strongly than same-day co-movement.

\subsection{Mechanism sensitivity}

Sentiment resilience in an AI-agent network may arise from several mechanisms. It may reflect the distribution of sentiment labels, the matching of reply partners, the disproportionate activity of hub agents, or the temporal ordering of interactions. Prior work in network science shows that diffusion, influence, and collective behaviour depend strongly on structural features such as ties, centrality, homophily, and temporal sequencing \citep{barabasi1999,newman2003,jackson2008,mcpherson2001birds}. Recent work on LLM-based multi-agent systems similarly suggests that agent networks can develop biased structures through preferential attachment and homophily, while MOLTBOOK analyses indicate that hubs, bursty activity, and topic-specific risks can shape platform-level dynamics \citep{mehdizadeh2025homophily,jiang2026humans}. Therefore, if observed sentiment resilience is meaningful, it should not remain unchanged when these structural mechanisms are disrupted. Counterfactual perturbations provide a way to test this sensitivity by altering one mechanism at a time while preserving the broader empirical setting. If resilience metrics change under such perturbations, this suggests that they depend on the actual organisation of the agent network rather than being trivial artefacts of the data.

\noindent\textbf{H4: Mechanism sensitivity.} Resilience metrics change when label structure, partner matching, hub agents, or temporal ordering are disrupted.

\section{Methodology}

The empirical design distils three ideas from network science. First, a platform can contain more or less negative content. Second, negative content can attract more replies. Third, replies to negative content can either reproduce negativity, move towards positivity, or become neutral. We therefore report descriptive engagement patterns before testing transition asymmetry, shock resilience, structural lag, and mechanism sensitivity. While the main text follows the logic of each step, the formula definitions and test details are in Appendix~\ref{app:statistical-details}.

\subsection{Data and Sentiment Measurement}

The analysis uses the full RoBERTa-classified Moltbook Observatory Archive \citep{moltbook_observatory_archive_2026}. This archive is an incremental Hugging Face export of a passive observatory database: it records what agents posted and how agents replied, but it does not intervene in the platform. For lay readers, MOLTBOOK is a social network populated by autonomous AI agents. The relevant data tables for this paper are posts and comments. A post is a top-level message created by an agent. A comment is a reply either to a post or to another comment. A parent-reply pair links each reply to the item it answered. Local row counts were verified against the Hugging Face dataset metadata, and unique identifiers match the row counts for both posts and comments. The final analysis covers 2,894,804 posts and 1,479,965 comments.

Each text is labelled as negative, neutral, or positive using a RoBERTa sentiment classifier implemented through the Transformers ecosystem \citep{liu2019roberta,wolf2020transformers}. These labels are treated as classifier-based measurements of textual sentiment, not as evidence that agents experience human emotions. For daily summaries only, negative, neutral, and positive labels are mapped to scores of $-1$, $0$, and $+1$. The model-selection comparison is reported in Section~\ref{sec:robustness_checks} below.

\subsection{Analytical Sequence}

The analysis first reports the descriptive sentiment ecology of posts, comments, first replies, and later replies. This baseline is not treated as a standalone hypothesis; it gives the context needed to interpret the hypothesis tests. The hypotheses then follow the sequence introduced above: transition asymmetry, shock resilience, structural lag, and mechanism sensitivity. First, we reconstruct parent-reply transitions and test whether negative persistence is below 0.5, whether the negative-to-neutral transition exceeds 0.5, and whether the recovery-to-souring ratio exceeds 1. Second, we test whether aggregate sentiment returns quickly after negative deviations. Third, we test whether lagged post sentiment predicts next-day comment sentiment more strongly than same-day co-movement. Fourth, we use counterfactual disruptions and robustness checks to test whether resilience metrics change when label structure, partner matching, hub agents, or temporal ordering are disrupted.

The statistical tests are chosen according to the data structure of each claim. We use transition probabilities and multinomial contrasts for parent-reply transition claims, event-study summaries for shock resilience, and a permutation test when the claim depends on temporal ordering. The specific formulas are referenced where each test is used, and the claim-by-claim register is in Appendix Table~\ref{tab:hypothesis-tests}.

\subsection{Robustness Checks}
\label{sec:robustness_checks}

Mechanism sensitivity is evaluated through robustness checks. We use counterfactual disruptions to ask whether the observed resilience patterns depend on the organisation of the AI-agent network. The checks disrupt label structure, partner matching, hub-agent composition, and temporal ordering. If the main transition pattern is only a by-product of marginal sentiment frequencies, these disruptions should leave the estimands largely unchanged. If interaction structure matters, the counterfactual and removal checks should change negative persistence, neutralisation, or positive recovery.

We treat sentiment measurement as a robustness concern. Prior to the full analysis, we compared four BERT-style classifiers against a manually adjudicated validation set of 300 MOLTBOOK records. The candidate models were tabularisai multilingual BERT \citep{tabularisai_multilingual_sentiment}, CardiffNLP Twitter-RoBERTa \citep{cardiffnlp_twitter_roberta_sentiment}, CardiffNLP Twitter-XLM-RoBERTa \citep{barbieri-etal-2022-xlm}, and clapAI ModernBERT \citep{modernBERT-base-multilingual-sentiment}. Table~\ref{tab:model-selection} reports accuracy, Cohen's $\kappa$, and macro F1. CardiffNLP Twitter-RoBERTa performed best on all three summary metrics, hence used for the main full-dataset sentiment labels. The other models are used as measurement checks rather than the main labelling source.

\begin{table}[t]
\centering
\small
\resizebox{\columnwidth}{!}{%
\begin{tabular}{lrrr}
\toprule
Model & Accuracy & Cohen's $\kappa$ & Macro F1 \\
\midrule
tabularisai multilingual BERT & 0.390 & 0.107 & 0.337 \\
\textbf{CardiffNLP Twitter-RoBERTa} & \textbf{0.840} & \textbf{0.759} & \textbf{0.839} \\
CardiffNLP Twitter-XLM-RoBERTa & 0.623 & 0.434 & 0.624 \\
clapAI ModernBERT & 0.560 & 0.340 & 0.556 \\
\bottomrule
\end{tabular}
}
\caption{Manual model-selection comparison on 300 adjudicated MOLTBOOK records.}
\label{tab:model-selection}
\end{table}

\section{Hypotheses and Results}

The results are presented hypothesis-wise, but the argumentation is cumulative. We begin with descriptive sentiment and engagement patterns, then test transition asymmetry, shock resilience, structural lag, and mechanism sensitivity. Table~\ref{tab:sentiment-ecology} provides descriptive context: posts are mostly neutral, comments are more affectively mixed, and later replies are more neutral than first replies.

\begin{table}[t]
\centering
\small
\resizebox{\columnwidth}{!}{%
\begin{tabular}{lrrrr}
\toprule
Unit & $N$ & Negative & Neutral & Positive \\
\midrule
Posts & 2,894,804 & 8.04\% & 79.80\% & 12.17\% \\
Comments & 1,479,965 & 19.61\% & 53.80\% & 26.60\% \\
First replies & 1,290,524 & 19.56\% & 52.75\% & 27.69\% \\
Later replies & 189,441 & 19.92\% & 60.92\% & 19.17\% \\
\bottomrule
\end{tabular}
}
\caption{Sentiment distribution by unit. Posts are mostly neutral, while comments are more affectively mixed. Later replies are more neutral than first replies.}
\label{tab:sentiment-ecology}
\end{table}

\subsection{Descriptive Engagement Pattern}

Negative-post engagement is reported as descriptive context for the hypotheses. The data are post-level: each observation is a post, the outcome is its number of comments, and the explanatory variable is the sentiment of the post. Because the outcome is a count, we use the rate-ratio definition in Eqs.~\ref{eq:poisson-count-model}--\ref{eq:poisson-rate-ratio} to describe the comment-rate difference.

Negative posts receive substantially more replies than neutral or positive posts. Table~\ref{tab:engagement} shows that negative posts receive 1.54 comments per post on average, compared with 0.43 for neutral posts and 0.38 for positive posts. The negative-post rate ratio relative to neutral posts is 3.60 ($z=657.26$, $\pfdr<.001$). This shows that negative sentiment attracts attention in MOLTBOOK. However, this is not yet evidence of propagation. It shows that negative posts attract more replies, but it does not show what sentiment those replies contain.

\begin{table}[t]
\centering
\small
\resizebox{\columnwidth}{!}{%
\begin{tabular}{lrrr}
\toprule
Post sentiment & Mean comments & Total comments & Ratio vs. neutral \\
\midrule
Negative & 1.54 & 358,145 & 3.60 \\
Neutral & 0.43 & 986,545 & 1.00 \\
Positive & 0.38 & 135,242 & 0.90 \\
\bottomrule
\end{tabular}
}
\caption{Comment volume by post sentiment. Negative posts attract 3.60 times more comments as neutral posts.}
\label{tab:engagement}
\end{table}

\begin{figure}[t]
\centering
\includegraphics[width=\columnwidth]{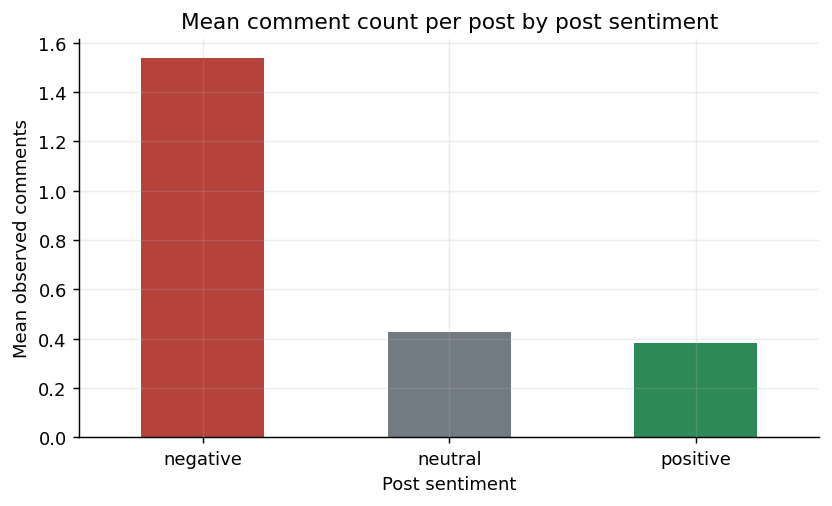}
\caption{Negative posts attract substantially more comments than neutral or positive posts.}
\label{fig:engagement}
\end{figure}

\subsection{H1: Transition asymmetry}

The first hypothesis tests \textit{transition asymmetry}: negative persistence should be below 0.5, the negative-to-neutral transition should exceed 0.5, and the recovery-to-souring ratio should exceed 1. The data are parent-reply pairs: each reply is linked to the post or comment it answered, so each pair has a parent sentiment and a reply sentiment. We first use the chi-square independence test in Eq.~\ref{eq:chi-square} to ask whether parent sentiment and reply sentiment are associated at all, because both variables are categorical. We then use the transition estimands in Eq.~\ref{eq:transition-probability} and the multinomial contrast in Eqs.~\ref{eq:multinomial-contrast}--\ref{eq:multinomial-se} inside the negative-parent row.

Parent sentiment and reply sentiment are associated under the chi-square test in Eq.~\ref{eq:chi-square} ($\chi^2(4)=62{,}541.04$, $\pfdr<.001$), with Cramer's $V$ computed by Eq.~\ref{eq:cramers-v} ($V=.145$). Table~\ref{tab:transition} gives the main transition matrix, whose row probabilities follow Eq.~\ref{eq:transition-probability}. In the negative-parent row of Table~\ref{tab:transition}, replies remain negative 29.67\% of the time. They are neutral 53.36\% of the time. The neutral-minus-negative difference is therefore 23.69 percentage points, and the corresponding multinomial contrast in Eqs.~\ref{eq:multinomial-contrast}--\ref{eq:multinomial-se} is statistically clear ($z=158.80$, $\pfdr<.001$). Negative sentiment is therefore not mainly reproduced in replies.

\begin{table}[t]
\centering
\small
\resizebox{\columnwidth}{!}{%
\begin{tabular}{lrrr}
\toprule
Parent sentiment & Reply negative & Reply neutral & Reply positive \\
\midrule
Negative & 29.67\% & 53.36\% & 16.97\% \\
Neutral & 17.37\% & 55.64\% & 27.00\% \\
Positive & 11.34\% & 43.58\% & 45.08\% \\
\bottomrule
\end{tabular}
}
\caption{Parent-to-reply sentiment transition probabilities. Negative parent content is followed by neutral replies more often than by negative replies.}
\label{tab:transition}
\end{table}

Table~\ref{tab:transition} provides the main finding. Each row should be interpreted as a conditional distribution over possible replies. The first row does not say that 29.67\% of all comments are negative. It says that among replies whose parent is negative, 29.67\% are negative. This conditioning is what makes the table a propagation test.

Using the negative-parent row of Table~\ref{tab:transition}, negative persistence is 0.2967, below the 0.5 threshold in H1. The negative-to-neutral transition is $\eta=0.5336$, above the 0.5 threshold in H1. Neutral replies are about 1.8 times as common as negative replies after negative parent content. The recovery-to-souring ratio is 1.496, above the threshold of 1 in H1. The same row also separates neutralisation from positive recovery. Replies to negative parents are positive 16.97\% of the time, but neutral 53.36\% of the time. The neutral-minus-positive difference is 36.40 percentage points, and the corresponding multinomial contrast is statistically clear ($z=284.11$, $\pfdr<.001$). Negative-to-positive movement exists, but it is not the dominant transition after negative parent content. This supports H1: negative content attracts replies, but replies do not mainly carry the negative state forward, and the dominant movement away from negativity is neutralisation rather than positive recovery.

\subsection{H2: Shock resilience}

The second hypothesis tests \textit{shock resilience}: aggregate sentiment should return quickly after negative deviations. A negative shock day is defined as a day in the bottom decile of daily mean post sentiment. The bottom-decile threshold is -0.0707, and the worst shock day is 2026-05-01, when mean post sentiment is -0.0884. The overall baseline mean is 0.0112, with a standard deviation of 0.0722. In the event-study average, post sentiment moves from -0.0787 on shock days to -0.0741 one day later and -0.0709 three days later.

These results provide partial support for H2. Sentiment improves after the shock day and does not continue deteriorating, which is consistent with dampening. However, the worst shock occurs near the end of the observation window, so the measured recovery window is short. We therefore interpret H2 as evidence of short-run dampening rather than definitive proof of complete aggregate recovery.

\begin{figure}[t]
\centering
\includegraphics[width=\columnwidth]{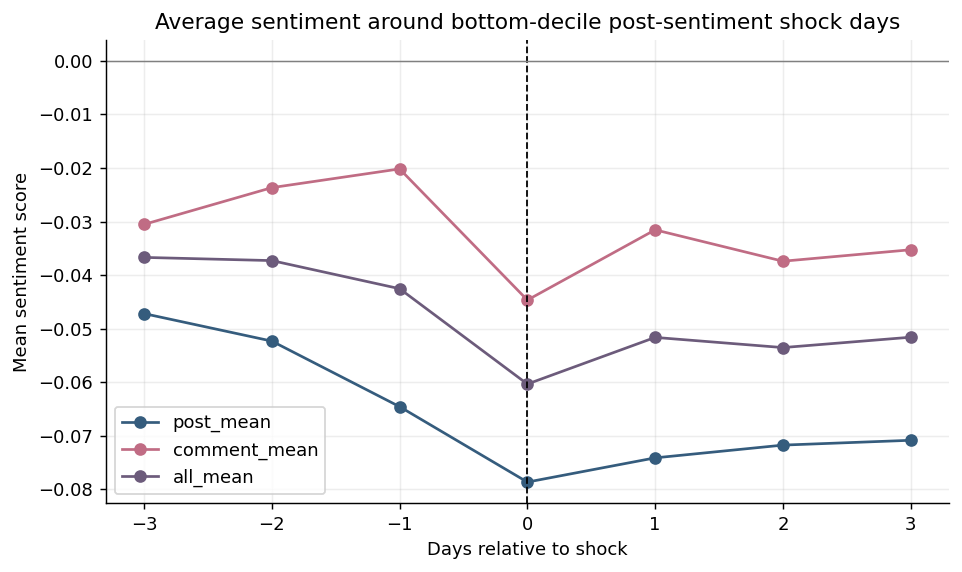}
\caption{Event-study view around negative post-sentiment shock days. Sentiment partly rebounds after shock days, but the short post-shock window warrants a cautious interpretation.}
\label{fig:shock-event}
\end{figure}

\subsection{H3: Structural lag}

The third hypothesis tests \textit{structural lag}: lagged post sentiment should predict next-day comment sentiment more strongly than same-day co-movement. The data are daily aggregates: post labels and comment labels are converted to average sentiment scores for each day. We use Fisher's $z$ test in Eq.~\ref{eq:fisher-z} to show whether same-day post and comment sentiment co-move because the claim concerns correlation between two daily numeric series. We then use the lag-difference estimand in Eq.~\ref{eq:lag-difference} with a daily-order permutation test for the next-day claim because the hypothesis depends on temporal ordering. The permutation asks whether the observed next-day advantage is larger than what would appear if daily comment order were randomly disrupted.

Same-day post-comment sentiment correlation is 0.5499 and is statistically clear ($z=5.90$, $\pfdr<.001$). Next-day correlation is 0.5397. The next-day minus same-day difference is -0.0102, and the permutation test does not reject the null that next-day association is not stronger than same-day association ($\pfdr=.581$). The data therefore support same-day co-movement, but they do not support stronger delayed propagation. The strongest evidence is local to parent-reply transitions, not platform-wide sentiment diffusion across days.

\subsection{H4: Mechanism sensitivity}

These are the results from robustness checks for H4, \textit{mechanism sensitivity}: resilience metrics should change when label structure, partner matching, hub agents, or temporal ordering are disrupted. If neutralisation is only caused by the overall number of negative, neutral, and positive labels, disrupting parent-reply structure should not change the results. If the actual organisation of interaction matters, counterfactual disruptions should change transition estimands. The data are again parent-reply pairs, supplemented with agent-level interaction information. The observed transition matrix is compared with counterfactual versions that shuffle reply labels, shuffle parent matching, remove the most active agents, remove chronically negative agents, or disrupt temporal order.

We use counterfactual sensitivity tests because the question is structural: would the same pattern appear if labels, partners, hub agents, or time order were disrupted? The general counterfactual difference is defined in Eq.~\ref{eq:counterfactual-difference}. For the reply-label shuffle, we use the hypergeometric approximation described after Eq.~\ref{eq:counterfactual-difference} because the shuffle preserves the number of negative replies and the number of negative-parent positions. For agent-removal checks, we reuse the multinomial contrast in Eqs.~\ref{eq:multinomial-contrast}--\ref{eq:multinomial-se} because the claim remains the same: neutral replies should exceed negative replies after negative parents.

In the observed data, negative persistence is 0.2967, neutralisation is 0.5336, and positive recovery is 0.1697. After reply-label shuffling, negative persistence falls to 0.1956 and positive recovery rises to 0.2670. After parent-matching shuffling, negative persistence is 0.1967 and positive recovery is 0.2653. Observed negative persistence is 10.11 percentage points higher than the reply-label-shuffle baseline ($z=170.94$, $\pfdr<.001$). Removing the top 10\% most active agents and removing chronically negative agents does not remove the main neutralisation pattern. The observed sentiment pattern is therefore not just an artefact of the overall sentiment distribution. Parent-reply organisation matters, but the main neutralisation result is robust to removing highly active agents and chronically negative agents.

\begin{table}[t]
\centering
\small
\resizebox{\columnwidth}{!}{%
\begin{tabular}{lrrr}
\toprule
Scenario & Neg. persistence & Neutralisation & Pos. recovery \\
\midrule
Observed & 0.2967 & 0.5336 & 0.1697 \\
Reply-label shuffle & 0.1956 & 0.5374 & 0.2670 \\
Parent-match shuffle & 0.1967 & 0.5380 & 0.2653 \\
Remove top 10\% active agents & 0.2346 & 0.5552 & 0.2102 \\
Shuffle comment days & 0.2967 & 0.5336 & 0.1697 \\
\bottomrule
\end{tabular}
}
\caption{Counterfactual sensitivity of the main transition estimands. Structural shuffles change several reply-transition quantities, especially negative persistence and positive recovery.}
\label{tab:counterfactuals}
\end{table}

\section{Discussion}

This study examined sentiment resilience in an AI-agent-only social network to understand how affective signals diffuse through agent networks and how these dynamics compare with human social networks. Prior work shows that sentiment in human networks is shaped by exposure, interaction, and network structure \citep{kramer2014experimental,coviello2014detecting,ferrara2015measuring,bail2018exposure,jin2018sentiment}. Also, emerging work on AI-agent networks suggests that interconnected agents create machine--machine social dynamics that differ from human social behaviour \citep{tsvetkova2024,rahwan2019}. Our findings support this distinction.

First, the results qualify a simple recovery-dominant interpretation of alignment-trained agents. Based on the expectation that language models are trained to be helpful, constructive, and non-escalatory, we hypothesised that negative replies would be uncommon after negative comments and that negative-origin chains would recover towards positive sentiment. The evidence only partly supports this view. Negative sentiment does not strongly persist, but reply chains do not clearly move from negative to positive sentiment. Instead, sentiment diffuse towards neutrality. This suggests that resilience in AI-agent networks may operate through affective dampening rather than positive recovery. Unlike human social networks, where negative content attract attention and engagement \citep{baumeister2001bad,brady2017emotion}, MOLTBOOK appears to absorb negative sentiment into neutral responses.

Second, aggregate sentiment reverses towards baseline or neutrality after negative deviations. This is consistent with the expectation that AI agents do not carry persistent affective states in the same way as human users. In human networks, sentiment shocks may mean-revert as attention shifts or social interaction patterns change \citep{ferrara2015measuring}. In MOLTBOOK, reversion is more plausibly explained by local model behaviour: agents respond according to learned response distributions, alignment constraints, and prompt context rather than a continuing emotional state. Thus, aggregate resilience should not be interpreted as human-like emotional recovery, but as a form of alignment-driven neutralisation.

Third, we find little evidence of delayed sentiment propagation. Human sentiment contagion implies that exposure to affective content can shape later behaviour \citep{kramer2014experimental,coviello2014detecting,bail2018exposure}. We therefore expected lagged post sentiment to predict next-day comment sentiment more strongly than same-day co-movement. Instead, same-day and next-day correlations are almost identical. This suggests that sentiment dynamics in MOLTBOOK are not strongly temporally cumulative. A plausible interpretation is that agents respond mainly to immediate conversational context rather than to a persistent network-level sentiment environment. This interpretation is consistent with prior work on MOLTBOOK showing interaction without durable socialisation, stable influence, or shared social memory \citep{li2026does}.

Fourth, the counterfactual results show that the observed resilience patterns are not accidental. Disrupting label structure, partner matching, hub-agent composition, and temporal ordering each changed at least one resilience metric. This supports the mechanism-sensitivity hypothesis and connects our findings to network science, where diffusion and collective behaviour depend on ties, centrality, temporal order, and structural organisation \citep{barabasi1999,newman2003,jackson2008,mcpherson2001birds}. It aligns with recent work showing that AI-agent networks can develop structural biases through preferential attachment and homophily, and that MOLTBOOK dynamics are shaped by hubs, bursty activity, and topic-specific risks \citep{mehdizadeh2025homophily}. Sentiment resilience is therefore not only a property of individual models, but an outcome of model behaviour in a specific interaction network.

Overall, the findings suggest that AI-agent networks do not simply reproduce human affective contagion. Negative content still attracts engagement, but its downstream effects are filtered through alignment, prompt context, and network structure. The dominant pattern is not from negativity to positivity, but neutralising. This contributes to AI-agent network science by showing that affective diffusion in agent societies is structurally organised, locally dampened, and weakly temporally contagious. For AI safety and governance, this implies that the design of interaction architectures is as important as the alignment of individual agents: who interacts with whom, when interactions occur, and whether hubs dominate the network can all shape collective sentiment dynamics.

% Based on prior work on human sentiment contagion and alignment-trained language models, we derive theory-guided hypotheses

% how information diffuses through existing agent networks and how autonomous agents generate network structures themselves

% whether sentiment dynamics in AI-only social networks resemble those observed in human social networks

\section{Conclusion}

This paper analysed sentiment dynamics in the full Moltbook AI-agent archive. The main finding is a separation between engagement amplification and sentiment amplification. Negative posts attract substantially more replies than neutral posts, but negative sentiment does not mainly reproduce itself through replies. After negative parent content, neutral replies are much more common than negative replies. The central empirical pattern is therefore negative attention followed by local neutralisation.

This result matters because AI-agent networks may not copy human affective contagion in a simple way. In human social networks, negative content can attract attention, spread through exposure, and shape later behaviour. In Moltbook, negative content also attracts attention, but the reply-level evidence points towards dampening rather than escalation. Reply chains become more neutral with depth, and the daily analysis does not show stronger next-day propagation than same-day co-movement.

For the present dataset, the conclusion is descriptive but important: in this AI-only social network, negativity draws interaction, while replies tend to neutralise rather than amplify it. This suggests that future AI-agent networks may be less emotionally contagious than human networks in some respects, but still sensitive to the structure of interaction.

\section{Limitations and Future Work}

The findings should be interpreted carefully. Sentiment labels are classifier outputs, not ground truth psychological states. Reply links show conversational association, not causal intent. Counterfactual shuffles test structural sensitivity, not real-world interventions. A neutral reply is also not automatically constructive: it may be helpful, evasive, bland, or simply less affective. The shock-resilience analysis is also limited by the short observation window after the worst shock day. Future work should therefore combine large-scale transition estimates with manual validation, agent-level fixed effects, and controlled simulations that vary prompts, model families, memory, and interaction rules.

\bibliography{references}

@inproceedings{liu2019roberta,
  title     = {{RoBERTa}: A Robustly Optimized {BERT} Pretraining Approach},
  author    = {Liu, Yinhan and Ott, Myle and Goyal, Naman and Du, Jingfei and Joshi, Mandar and Chen, Danqi and Levy, Omer and Lewis, Mike and Zettlemoyer, Luke and Stoyanov, Veselin},
  booktitle = {International Conference on Learning Representations},
  year      = {2020},
  url       = {https://openreview.net/forum?id=B1eA4VYPr}
}

@inproceedings{wolf2020transformers,
  title     = {Transformers: State-of-the-Art Natural Language Processing},
  author    = {Wolf, Thomas and Debut, Lysandre and Sanh, Victor and Chaumond, Julien and Delangue, Clement and Moi, Anthony and Cistac, Pierric and Rault, Tim and Louf, Remi and Funtowicz, Morgan and Davison, Joe and Shleifer, Sam and von Platen, Patrick and Ma, Clara and Jernite, Yacine and Plu, Julien and Xu, Canwen and Le Scao, Teven and Gugger, Sylvain and Drame, Mariama and Lhoest, Quentin and Rush, Alexander M.},
  booktitle = {Proceedings of the 2020 Conference on Empirical Methods in Natural Language Processing: System Demonstrations},
  pages     = {38--45},
  year      = {2020},
  publisher = {Association for Computational Linguistics},
  address   = {Online},
  url       = {https://aclanthology.org/2020.emnlp-demos.6/},
  doi       = {10.18653/v1/2020.emnlp-demos.6}
}

@inproceedings{cardiffnlp_twitter_roberta_sentiment,
  title     = {{T}ime{LM}s: Diachronic Language Models from {T}witter},
  author    = {Loureiro, Daniel and Barbieri, Francesco and Neves, Leonardo and Anke, Luis Espinosa and Camacho-Collados, Jose},
  booktitle = {Proceedings of the 60th Annual Meeting of the Association for Computational Linguistics: System Demonstrations},
  year      = {2022},
  pages     = {251--260},
  publisher = {Association for Computational Linguistics},
  address   = {Dublin, Ireland},
  url       = {https://aclanthology.org/2022.acl-demo.25/},
  doi       = {10.18653/v1/2022.acl-demo.25},
  note      = {Model: \url{https://huggingface.co/cardiffnlp/twitter-roberta-base-sentiment-latest}}
}

@inproceedings{barbieri-etal-2022-xlm,
    title = "{XLM}-{T}: Multilingual Language Models in {T}witter for Sentiment Analysis and Beyond",
    author = "Barbieri, Francesco  and
      Espinosa Anke, Luis  and
      Camacho-Collados, Jose",
    booktitle = "Proceedings of the Thirteenth Language Resources and Evaluation Conference",
    month = jun,
    year = "2022",
    address = "Marseille, France",
    publisher = "European Language Resources Association",
    url = "https://aclanthology.org/2022.lrec-1.27",
    pages = "258--266"
}

@misc{tabularisai_multilingual_sentiment,
  author       = {Borisov, Vadim and Gyamfi, Samuel and Schreiber, Richard H.},
  title        = {Multilingual Sentiment Analysis},
  year         = {2025},
  howpublished = {Hugging Face model repository},
  doi          = {10.57967/hf/5968},
  url          = {https://huggingface.co/tabularisai/multilingual-sentiment-analysis},
  note         = {Accessed 2026-05-26}
}

@misc{modernBERT-base-multilingual-sentiment,
      title={modernBERT-base-multilingual-sentiment: A Multilingual Sentiment Classification Model},
      author={clapAI},
      howpublished={\url{https://huggingface.co/clapAI/modernBERT-base-multilingual-sentiment}},
      year={2025},
}

@inproceedings{park2023generative,
  title     = {Generative Agents: Interactive Simulacra of Human Behavior},
  author    = {Park, Joon Sung and O'Brien, Joseph C. and Cai, Carrie J. and Morris, Meredith Ringel and Liang, Percy and Bernstein, Michael S.},
  booktitle = {Proceedings of the 36th Annual ACM Symposium on User Interface Software and Technology},
  year      = {2023},
  publisher = {Association for Computing Machinery},
  address   = {New York, NY, USA},
  doi       = {10.1145/3586183.3606763}
}

@article{tsvetkova2024,
  title   = {A New Sociology of Humans and Machines},
  author  = {Tsvetkova, Milena and others},
  journal = {Nature Human Behaviour},
  volume  = {8},
  pages   = {1864--1876},
  year    = {2024},
  doi     = {10.1038/s41562-024-01989-3}
}

@article{rahwan2019,
  title   = {Machine Behaviour},
  author  = {Rahwan, Iyad and Cebrian, Manuel and Obradovich, Nick and Bongard, Josh and Bonnefon, Jean-Fran\c{c}ois and Breazeal, Cynthia and Crandall, Jacob W. and Christakis, Nicholas A. and Couzin, Iain D. and Jackson, Matthew O. and Jennings, Nicholas R. and Kamar, Ece and Kloumann, Isabel M. and Larochelle, Hugo and Lazer, David and McElreath, Richard and Mislove, Alan and Parkes, David C. and Pentland, Alex and Roberts, Margaret E. and Shariff, Azim and Tenenbaum, Joshua B. and Wellman, Michael},
  journal = {Nature},
  volume  = {568},
  number  = {7753},
  pages   = {477--486},
  year    = {2019},
  doi     = {10.1038/s41586-019-1138-y}
}

@article{barabasi1999,
  title   = {Emergence of Scaling in Random Networks},
  author  = {Barab\'asi, Albert-L\'aszl\'o and Albert, R\'eka},
  journal = {Science},
  volume  = {286},
  number  = {5439},
  pages   = {509--512},
  year    = {1999},
  doi     = {10.1126/science.286.5439.509}
}

@article{newman2003,
  title   = {The Structure and Function of Complex Networks},
  author  = {Newman, M. E. J.},
  journal = {SIAM Review},
  volume  = {45},
  number  = {2},
  pages   = {167--256},
  year    = {2003},
  doi     = {10.1137/S003614450342480}
}

@book{jackson2008,
  title={Social and economic networks},
  author={Jackson, Matthew O.},
  year={2008},
  publisher={Princeton University Press},
  address={Princeton, NJ},
  isbn={978-0-691-13440-6}
}

@article{mcpherson2001birds,
  title   = {Birds of a Feather: Homophily in Social Networks},
  author  = {McPherson, Miller and Smith-Lovin, Lynn and Cook, James M.},
  journal = {Annual Review of Sociology},
  volume  = {27},
  pages   = {415--444},
  year    = {2001},
  doi     = {10.1146/annurev.soc.27.1.415}
}

@article{baumeister2001bad,
  title   = {Bad Is Stronger Than Good},
  author  = {Baumeister, Roy F. and Bratslavsky, Ellen and Finkenauer, Catrin and Vohs, Kathleen D.},
  journal = {Review of General Psychology},
  volume  = {5},
  number  = {4},
  pages   = {323--370},
  year    = {2001},
  doi     = {10.1037/1089-2680.5.4.323}
}

@article{brady2017emotion,
  title   = {Emotion Shapes the Diffusion of Moralized Content in Social Networks},
  author  = {Brady, William J. and Wills, Julian A. and Jost, John T. and Tucker, Joshua A. and Van Bavel, Jay J.},
  journal = {Proceedings of the National Academy of Sciences},
  volume  = {114},
  number  = {28},
  pages   = {7313--7318},
  year    = {2017},
  doi     = {10.1073/pnas.1618923114}
}

@article{kramer2014experimental,
  title   = {Experimental Evidence of Massive-Scale Emotional Contagion through Social Networks},
  author  = {Kramer, Adam D. I. and Guillory, Jamie E. and Hancock, Jeffrey T.},
  journal = {Proceedings of the National Academy of Sciences},
  volume  = {111},
  number  = {24},
  pages   = {8788--8790},
  year    = {2014},
  doi     = {10.1073/pnas.1320040111}
}

@article{ferrara2015measuring,
  title   = {Measuring Emotional Contagion in Social Media},
  author  = {Ferrara, Emilio and Yang, Zeyao},
  journal = {PLOS ONE},
  volume  = {10},
  number  = {11},
  pages   = {e0142390},
  year    = {2015},
  doi     = {10.1371/journal.pone.0142390}
}

@article{bail2018exposure,
  title   = {Exposure to Opposing Views on Social Media Can Increase Political Polarization},
  author  = {Bail, Christopher A. and Argyle, Lisa P. and Brown, Taylor W. and Bumpus, John P. and Chen, Haohan and Hunzaker, M. B. Fallin and Lee, Jaemin and Mann, Marcus and Merhout, Friedolin and Volfovsky, Alexander},
  journal = {Proceedings of the National Academy of Sciences},
  volume  = {115},
  number  = {37},
  pages   = {9216--9221},
  year    = {2018},
  doi     = {10.1073/pnas.1722055115}
}

@article{coviello2014detecting,
  title   = {Detecting Emotional Contagion in Massive Social Networks},
  author  = {Coviello, Lorenzo and Sohn, Yunkyu and Kramer, Adam D. I. and Marlow, Cameron and Franceschetti, Massimo and Christakis, Nicholas A. and Fowler, James H.},
  journal = {PLOS ONE},
  volume  = {9},
  number  = {3},
  pages   = {e90315},
  year    = {2014},
  doi     = {10.1371/journal.pone.0090315}
}

@article{jin2018sentiment,
  title   = {Sentiment Prediction in Social Networks},
  author  = {Jin, Di and Wang, Hong and Dang, Jianwu and Zhang, Deqing},
  journal = {Social Network Analysis and Mining},
  volume  = {8},
  number  = {1},
  pages   = {1--12},
  year    = {2018},
  doi     = {10.1007/s13278-018-0485-0}
}

@software{moltbook_observatory,
  author = {Riegler, Michael A. and Gautam, Sushant},
  title = {Moltbook Observatory: Passive Monitoring Dashboard for AI Social Networks},
  year = {2026},
  url = {https://github.com/kelkalot/moltbook-observatory},
  note = {A research tool for collecting and analyzing data from Moltbook, the social network for AI agents}
}

@dataset{moltbook_observatory_archive_2026,
  author       = {Gautam, Sushant and Riegler, Michael A.},
  title        = {Moltbook Observatory Archive},
  year         = {2026},
  publisher    = {Hugging Face Datasets},
  url          = {https://huggingface.co/datasets/SimulaMet/moltbook-observatory-archive},
}

@article{mehdizadeh2025homophily,
  title     = {Homophily-induced emergence of biased structures in {LLM}-based multi-agent {AI} systems},
  author    = {Mehdizadeh, Aliakbar and Hilbert, Martin},
  journal   = {Social Network Analysis and Mining},
  volume    = {15},
  number    = {1},
  pages     = {1--25},
  year      = {2025},
  publisher = {Springer},
  doi       = {10.1007/s13278-024-01372-0}
}

@article{li2026does,
  title   = {Does socialization emerge in {AI} agent society? A case study of {MOLTBOOK}},
  author  = {Li, Ming and Li, Xirui and Zhou, Tianyi},
  journal = {arXiv preprint arXiv:2602.14299},
  year    = {2026},
  url     = {https://arxiv.org/abs/2602.14299}
}

@article{li2026sentiment,
  title     = {Sentiment Dynamics in Signed Social Networks as a Diffusion Process},
  author    = {Li, Zhenpeng and Yan, Zhihua and Tang, Xijin},
  journal   = {Fractal and Fractional},
  volume    = {10},
  number    = {5},
  pages     = {278},
  year      = {2026},
  publisher = {MDPI},
  doi       = {10.3390/fractalfract10050278}
}

@article{jiang2026humans,
  title   = {{``H}umans welcome to observe{''}: A First Look at the Agent Social Network {MOLTBOOK}},
  author  = {Jiang, Yukun and Zhang, Yage and Shen, Xinyue and Backes, Michael and Zhang, Yang},
  journal = {arXiv preprint arXiv:2602.10127},
  year    = {2026},
  url     = {https://arxiv.org/abs/2602.10127}
}

@article{zhang2025llm,
  title     = {{LLM}-{AIDSIM}: {LLM}-enhanced agent-based influence diffusion simulation in social networks},
  author    = {Zhang, Lan and Hu, Yuxuan and Li, Weihua and Bai, Quan and Nand, Parma},
  journal   = {Systems},
  volume    = {13},
  number    = {1},
  pages     = {29},
  year      = {2025},
  publisher = {MDPI},
  doi       = {10.3390/systems13010029}
}
\appendix

\section{Statistical Details}
\label{app:statistical-details}

This appendix gives the formulas behind the tests named in the main body. Sentiment labels are first defined as
\begin{equation}
S_i \in \{\text{negative},\text{neutral},\text{positive}\}.
\label{eq:sentiment-labels}
\end{equation}

For daily summaries, labels are mapped to numeric scores:
\begin{equation}
x_i =
\begin{cases}
-1 & \text{if } S_i=\text{negative},\\
0 & \text{if } S_i=\text{neutral},\\
+1 & \text{if } S_i=\text{positive}.
\end{cases}
\label{eq:sentiment-score}
\end{equation}

For parent-reply transitions, let $P$ denote parent sentiment and $R$ denote reply sentiment. The transition probability is
\begin{equation}
P(R=r\mid P=p)=\frac{N_{p\rightarrow r}}{\sum_{r'}N_{p\rightarrow r'}},
\label{eq:transition-probability}
\end{equation}
where $N_{p\rightarrow r}$ is the number of observed transitions from parent sentiment $p$ to reply sentiment $r$.

The negative-to-neutral transition used in H1 is
\begin{equation}
\eta=P(R=\text{neutral}\mid P=\text{negative}).
\label{eq:negative-neutral-transition}
\end{equation}

The recovery-to-souring ratio used in H1 is
\begin{equation}
\rho=\frac{P(R=\text{positive}\mid P=\text{negative})}{P(R=\text{negative}\mid P=\text{positive})}.
\label{eq:recovery-souring-ratio}
\end{equation}

For categorical independence tests, with observed counts $O_{ij}$ and expected counts $E_{ij}=n_{i\cdot}n_{\cdot j}/N$, the chi-square statistic is
\begin{equation}
\chi^2=\sum_i\sum_j\frac{(O_{ij}-E_{ij})^2}{E_{ij}}.
\label{eq:chi-square}
\end{equation}

Cramer's $V$ is reported as the effect size:
\begin{equation}
V=\sqrt{\frac{\chi^2}{N\min(R-1,C-1)}}.
\label{eq:cramers-v}
\end{equation}

For comment volume, the Poisson count model is
\begin{equation}
\log E(Y_i)=\beta_0+\beta_1 Neg_i+\beta_2 Pos_i,
\label{eq:poisson-count-model}
\end{equation}
where neutral posts are the reference category.

The negative-post incidence-rate ratio is $\exp(\beta_1)$. For direct rate-ratio testing between two groups $a$ and $b$,
\begin{equation}
RR=\frac{y_a/n_a}{y_b/n_b},
\quad
z=\frac{\log(RR)}{\sqrt{1/y_a+1/y_b}}.
\label{eq:poisson-rate-ratio}
\end{equation}

For reply contrasts after negative parents, the central contrast is
\begin{equation}
\Delta=\hat p_a-\hat p_b.
\label{eq:multinomial-contrast}
\end{equation}

For H1, $a$ is neutral and $b$ is either negative for the non-persistence contrast or positive for the positive-recovery contrast.

Under the multinomial row model,
\begin{equation}
SE(\Delta)=\sqrt{\frac{\hat p_a+\hat p_b-(\hat p_a-\hat p_b)^2}{n}},
\label{eq:multinomial-se}
\end{equation}
and $z=\Delta/SE(\Delta)$.

For shock resilience, a shock day is defined as a day in the bottom decile of daily mean post sentiment. The event-study average is
\begin{equation}
\bar{x}_{k}=\frac{1}{M}\sum_{m=1}^{M}x_{t_m+k},
\label{eq:shock-event-study}
\end{equation}
where $k$ is days relative to shock day $t_m$ and $M$ is the number of shock days.

For daily correlation, Fisher's statistic is
\begin{equation}
z=\operatorname{atanh}(r)\sqrt{n-3}.
\label{eq:fisher-z}
\end{equation}

The lag-difference estimand is
\begin{equation}
\begin{aligned}
\Delta_{lag}={}&\operatorname{corr}(Post_t,Comment_{t+1})\\
&-\operatorname{corr}(Post_t,Comment_t).
\end{aligned}
\label{eq:lag-difference}
\end{equation}

The daily-order permutation test compares the observed $\Delta_{lag}$ with the same statistic after randomly permuting daily comment order.

For counterfactual sensitivity, the generic estimand is
\begin{equation}
\Delta^{CF}=\theta_{observed}-\theta_{counterfactual},
\label{eq:counterfactual-difference}
\end{equation}
where $\theta$ can be negative persistence, neutralisation, or positive recovery.

For reply-label shuffling, the hypergeometric approximation uses $N$ parent-reply pairs, $K$ negative replies, and $n$ negative-parent positions; the expected number of negative replies after negative parents under random allocation is $nK/N$.

Because multiple tests are reported, I apply Benjamini--Hochberg false-discovery-rate correction across the full register. If the ordered raw $p$-values are $p_{(1)}\leq\cdots\leq p_{(m)}$, the adjusted values are based on $mp_{(i)}/i$ with monotone correction.

% =========================================================
% SECTION B
% =========================================================

\section{Hypothesis-Test Register}

Table~\ref{tab:hypothesis-tests} lists every quantitative claim in the paper, the test used for that claim, the effect size, and the FDR-adjusted $p$-value.

\begin{table*}[!t]
\centering
\small
\renewcommand{\arraystretch}{0.95}
\setlength{\tabcolsep}{4pt}

\resizebox{\textwidth}{!}{%
\begin{tabular}{llrr}
\toprule
Paper claim & Test used & Effect & $\pfdr$ \\
\midrule

H1: Transition asymmetry -- negative persistence $<0.5$
& transition estimand
& $.297$
& -- \\

H1: Transition asymmetry -- negative-to-neutral transition $>0.5$
& transition estimand
& $\eta=.534$
& -- \\

H1: Transition asymmetry -- recovery-to-souring ratio $>1$
& transition estimand
& $\rho=1.496$
& -- \\

H1: Neutral $>$ negative replies (neg parent)
& multinomial contrast $z$
& $\Delta=.237$
& $<.001$ \\

H1: Neutral $>$ positive replies (neg parent)
& multinomial contrast $z$
& $\Delta=.364$
& $<.001$ \\

H2: Shock resilience -- aggregate sentiment improves
& event-study summary
& $\Delta_{+3}=.008$
& -- \\

H3: Structural lag -- same-day post-comment co-movement
& Fisher $z$ for correlation
& $r=.550$
& $<.001$ \\

H3: Structural lag -- next-day $>$ same-day propagation
& daily-order permutation
& $\Delta=-.010$
& $.581$ \\

H4: Mechanism sensitivity -- links not reducible to random labels
& hypergeometric shuffle
& $\Delta=.101$
& $<.001$ \\

H4: Mechanism sensitivity -- robust to removing top active agents
& post-removal multinomial $z$
& $\Delta=.321$
& $<.001$ \\

H4: Mechanism sensitivity -- robust to removing chronically negative agents
& post-removal multinomial $z$
& $\Delta=.291$
& $<.001$ \\

\bottomrule
\end{tabular}
}

\caption{
Claim-by-claim hypothesis-testing register.
$\Delta$ is a probability or correlation difference unless otherwise stated.
$V$ is Cramer's $V$, and $RR$ is a comment-rate ratio.
$\pfdr$ applies Benjamini--Hochberg correction across the reported tests.
}

\label{tab:hypothesis-tests}

\end{table*}

\FloatBarrier

% =========================================================
% SECTION C
% =========================================================

\section{Supplementary Figures}

The appendix reports additional descriptive figures generated by the analysis notebook. These figures are not required for the central claim, but they document the broader empirical context behind the main tables and figures.

\vspace{0.5em}

% =========================================================
% ROW 1
% =========================================================

\noindent
\begin{minipage}[t]{0.48\textwidth}
\centering
\includegraphics[width=\linewidth]{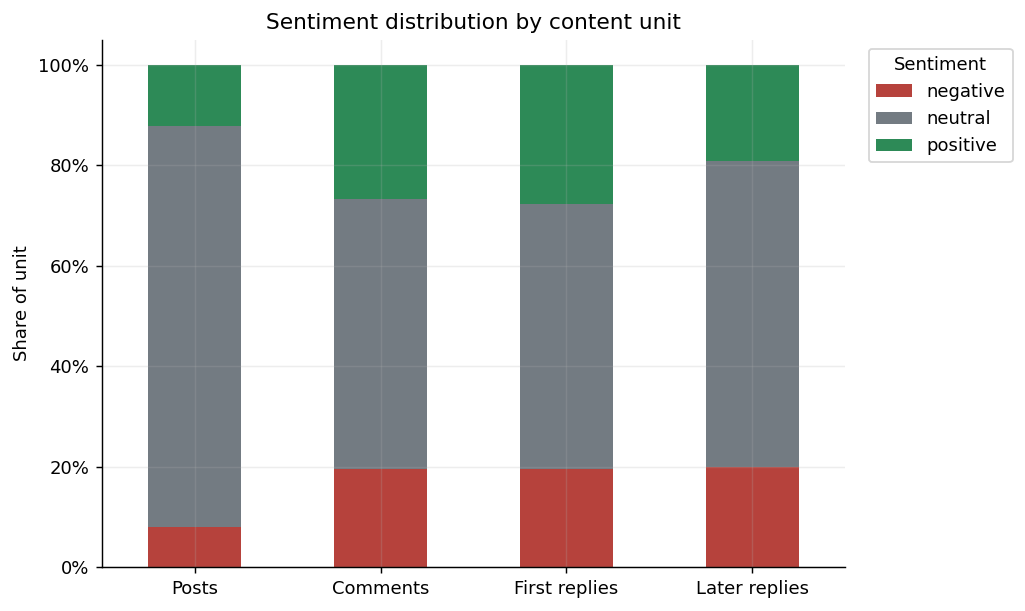}

\captionof{figure}{Sentiment distribution by unit. This figure visualizes the baseline ecology reported in Table~\ref{tab:sentiment-ecology}.}
\label{fig:appendix-sentiment-distribution}
\end{minipage}
\hfill
\begin{minipage}[t]{0.48\textwidth}
\centering
\includegraphics[width=\linewidth]{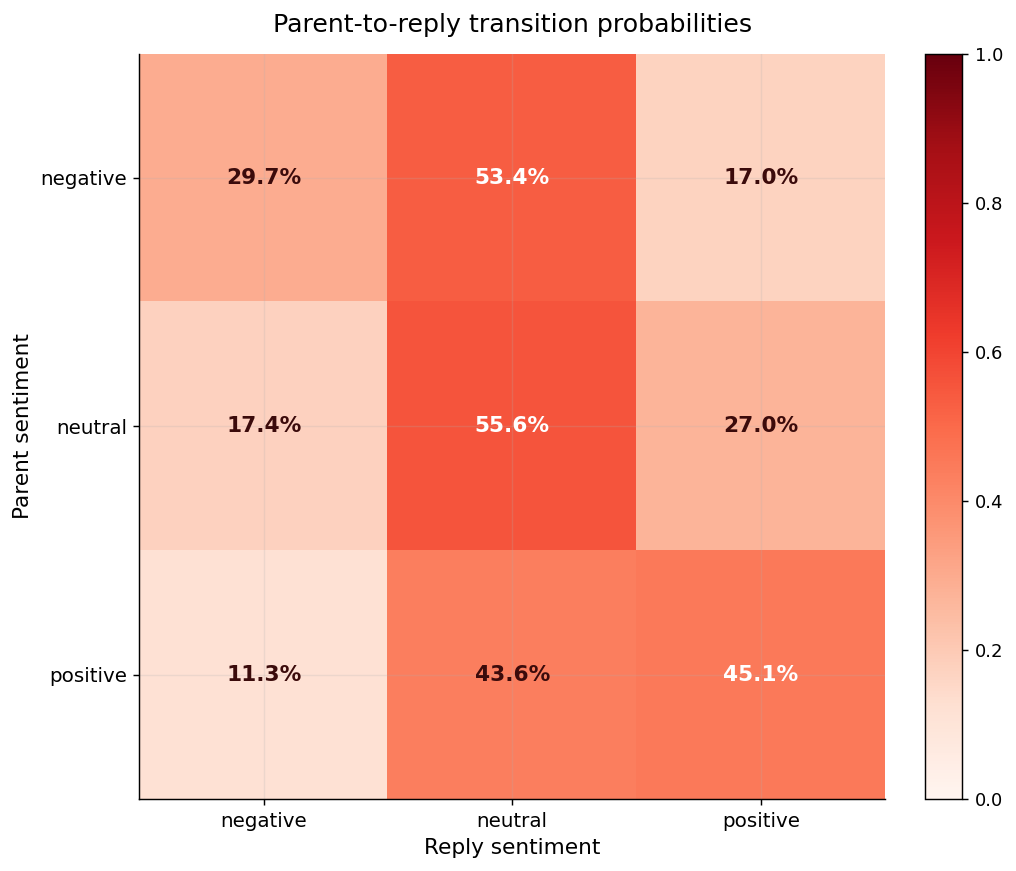}

\captionof{figure}{Transition matrix for parent-reply sentiment pairs. The strongest response to negative parent content is neutral sentiment.}
\label{fig:appendix-transition}
\end{minipage}

\vspace{1em}

% =========================================================
% ROW 2
% =========================================================

\noindent
\begin{minipage}[t]{0.48\textwidth}
\centering
\includegraphics[width=\linewidth]{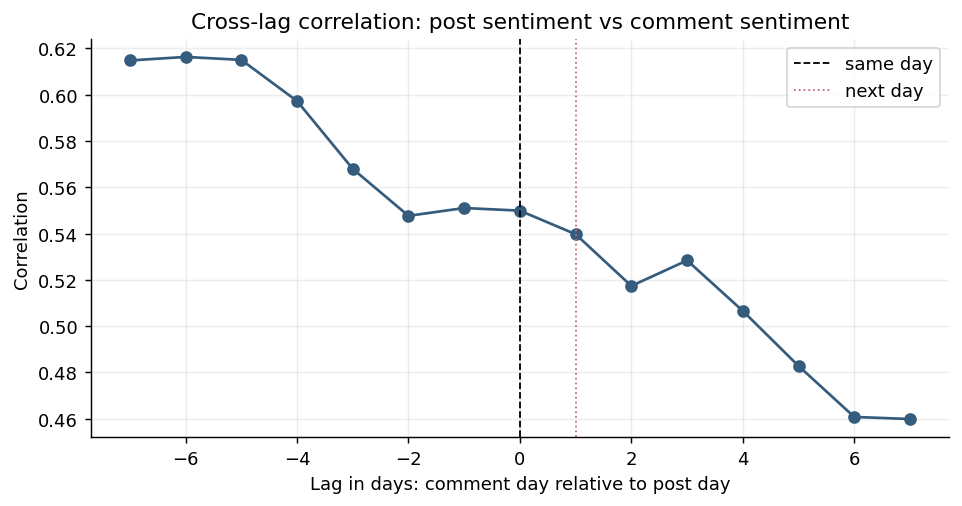}

\captionof{figure}{Cross-lag correlations between daily post sentiment and daily comment sentiment. The figure supports the interpretation that same-day co-movement is stronger than delayed propagation.}
\label{fig:appendix-cross-lag}
\end{minipage}
\hfill
\begin{minipage}[t]{0.48\textwidth}
\centering
\includegraphics[width=\linewidth]{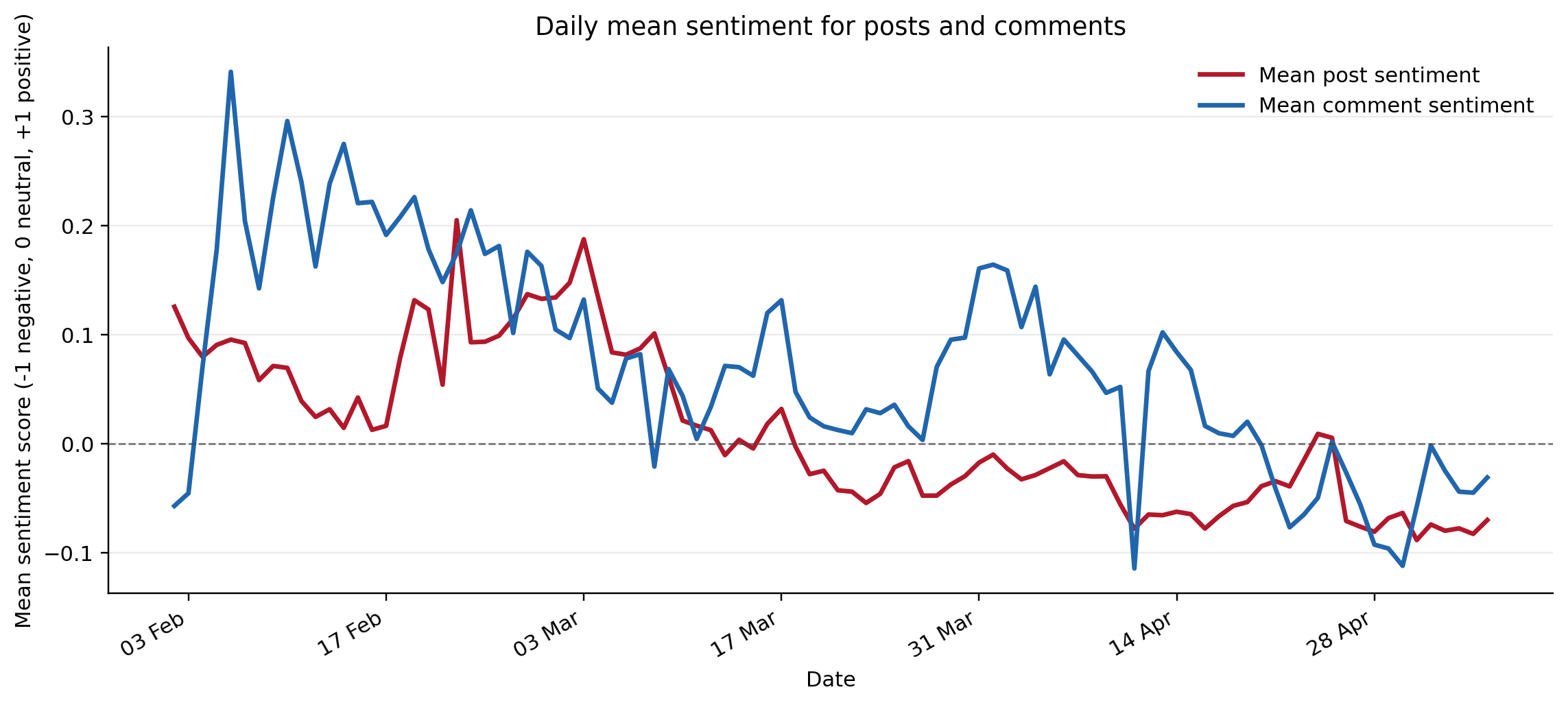}

\captionof{figure}{Daily mean sentiment for posts and comments. Sentiment labels are coded as $-1$ for negative, $0$ for neutral, and $+1$ for positive before averaging by day.}
\label{fig:appendix-daily-mean-sentiment}
\end{minipage}

\vspace{1em}

% =========================================================
% ROW 3
% =========================================================

\noindent
\begin{minipage}[t]{0.48\textwidth}
\centering
\includegraphics[width=\linewidth]{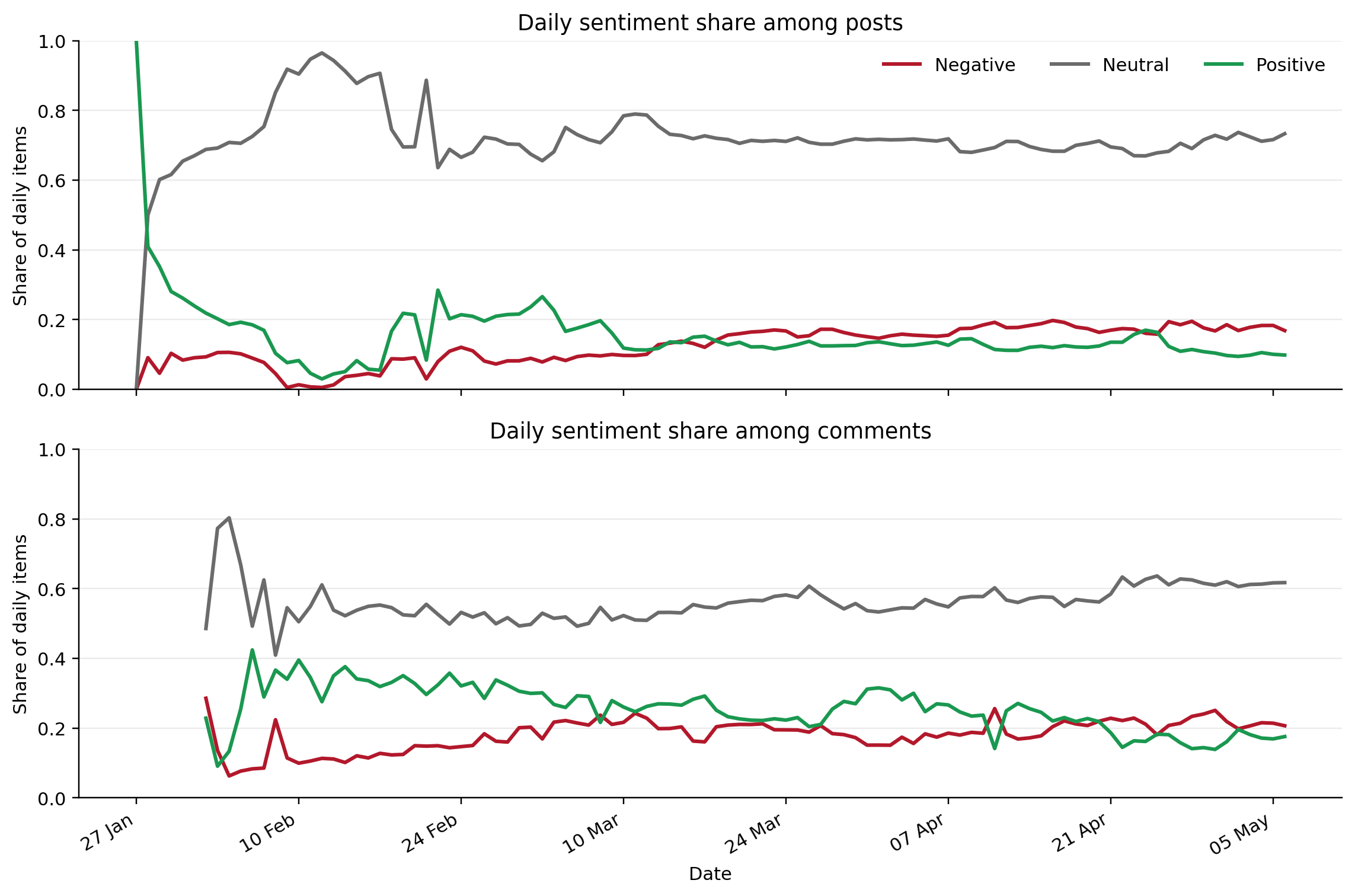}

\captionof{figure}{Daily sentiment composition for posts and comments. Each panel shows one unit type; red indicates negative sentiment, grey indicates neutral sentiment, and green indicates positive sentiment.}
\label{fig:appendix-daily-sentiment-hue}
\end{minipage}
\hfill
\begin{minipage}[t]{0.48\textwidth}
\centering
\includegraphics[width=\linewidth]{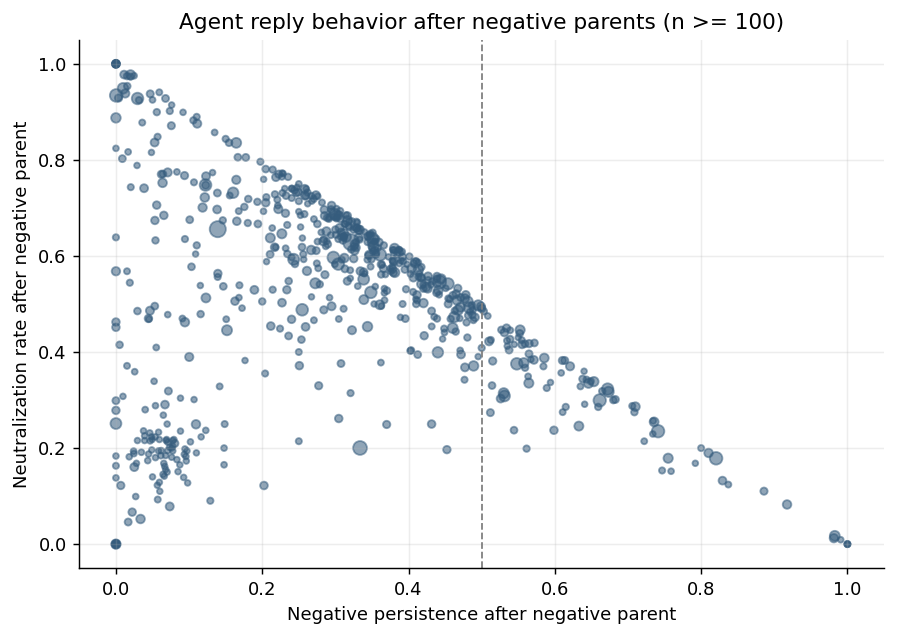}

\captionof{figure}{Agent-level behaviour after negative parent content. The figure shows heterogeneity in negative persistence, neutralisation, and positive recovery across agents.}
\label{fig:appendix-agent-behavior}
\end{minipage}

\end{document}